\documentclass{BNM-zoubs}

\begin{document}

\markboth{Bing-Song Zou and Ju-Jun Xie}{On $pp \to p K \Lambda, N K \Sigma, pp \phi$ \\
--- the basic ingredients for strangeness production in heavy ion
collisions}

\catchline{}{}{}{}{}

\title{On $pp \to p K \Lambda, N K \Sigma, pp \phi$ \\
--- the basic ingredients for strangeness production in heavy ion
collisions}

\author{Bing-Song Zou and Ju-Jun Xie}

\address{Institute of High Energy Physics, Chinese Academy of
sciences, Beijing 100049, China\\
Theoretical Physics Center for Science Facilities, CAS, Beijing
100049, China\\
zoubs@mail.ihep.ac.cn; \quad
xiejujun@mail.ihep.ac.cn}

\maketitle

\begin{history}
\received{(received date)}
\revised{(revised date)}
\end{history}

\begin{abstract}
The strangeness production in heavy ion collisions was proposed to
be probes of the nuclear equation of state, Kaon potential in
nuclear medium, strange quark matter and quark-gluon plasma, etc.
However, to act as reliable probes, proper understanding of the
basic ingredients for the strangeness production, such as $pp \to
pK^+\Lambda$, $pp \to pp \phi$ and $pp \to nK^+\Sigma^+$ is
necessary. Recent study of these reactions clearly shows that
previously ignored contributions from the spin-parity $1/2^-$
resonances, $N^*(1535)$ and $\Delta^*(1620)$, are in fact very
important for these reactions, especially for near-threshold
energies. It is necessary to include these contributions for getting
reliable calculation for the strangeness production in heavy ion
collisions.
\end{abstract}

\section{Introduction}
Strangeness production in heavy ion collisions is presently an issue
of intense study since it plays important roles in many aspects.
Because $K^+$ mesons have a long mean free path inside the nuclear
matter, they are believed to be good messengers to provide
information about the high density and temperature phase of the
heavy ion collisions~\cite{schne,baker}. The subthreshold Kaon
production was proposed to be a sensitive probe of the nuclear
equation of state (EOS)~\cite{aiche} while Kaon flow was proposed to
be a probe of the Kaon potential in nuclear medium~\cite{ligq}. The
strangeness production was also proposed to be good probe of
possible formation of quark-gluon-plasma (QGP) \cite{greiner,shor}.
Especially, in Ref.~\cite{shor}, Shor suggested the $\phi$ meson
production to be an ideal candidate to study the QGP in nuclear
collisions, due to its flavor contents composed of a strange and
antistrange quark.

\begin{figure}[th]
\centerline{\psfig{file=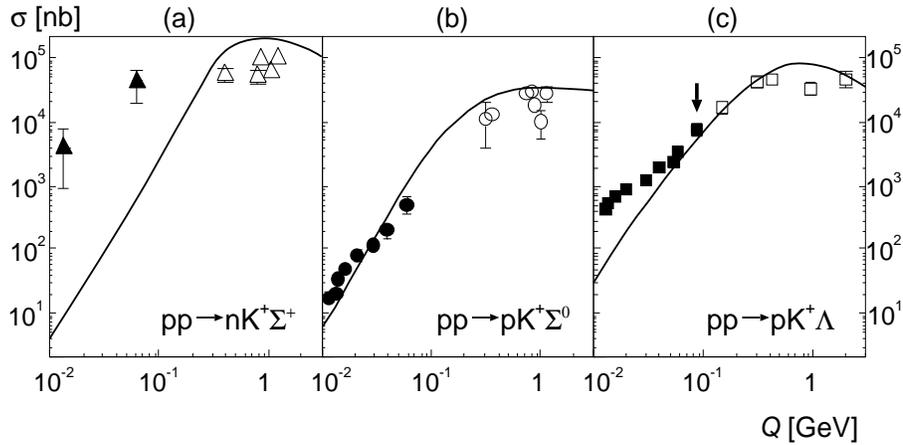,height=6.0cm,width=12cm}}
\vspace*{0pt} \caption{Total cross section for various $pp \to N K^+
Y$ reactions$^7$. Data in the near-threshold region (full symbols)
are from recent COSY experiments$^{7,8,9}$ and data at high excess
energies (open symbols) are from older experiments$^{10}$. Solid
curves are the resonance model predictions$^{11,12}$.} \label{ktcs}
\end{figure}

To act as reliable probes, proper understanding of the basic
ingredients for the strangeness production, such as $pp \to
pK^+\Lambda$, $pp \to pp \phi$ and $pp \to nK^+\Sigma^+$, is
necessary. Status for various $pp \to N K^+ Y$
reactions~\cite{rozek} is shown by Fig.\ref{ktcs}. While a typical
resonance model \cite{sibir99,tsushima99} fits the older data
\cite{baldi} at high excess energies quite well, it underestimates
recent COSY data at near-threshold energies for $pp\to nK^+\Sigma^+$
and $pp\to pK^+\Lambda$ by order(s) of magnitude. Other model
calculations \cite{linpa,faldt,tsushima,gaspar} suffer similar
problem for their predictions at near-threshold energies. The
situation for $pp \to pp \phi$ is even worse. The data is scarce due
to much smaller cross section. Only recently some near-threshold
data appeared \cite{balestra,hartman} with little theoretical study
available.

Recently these reactions have been restudied by including
contributions from previously ignored subthreshold $N^*(1535)$ and
$\Delta^*(1620)$ resonances \cite{liubc,liubc2,xieplb,xiephi}. The
fit to the data are much improved. Although the original motivation
for studying these reactions is to improve our understanding of the
internal quark structure of relevant baryons~\cite{zou}, the results
turn out to be also very important for studying strangeness
production in heavy ion collisions. Hence we summarize main results
from these studies here.

\section{Study on $pp \to pK^+\Lambda$ reaction}

\begin{figure}
\psfig{file=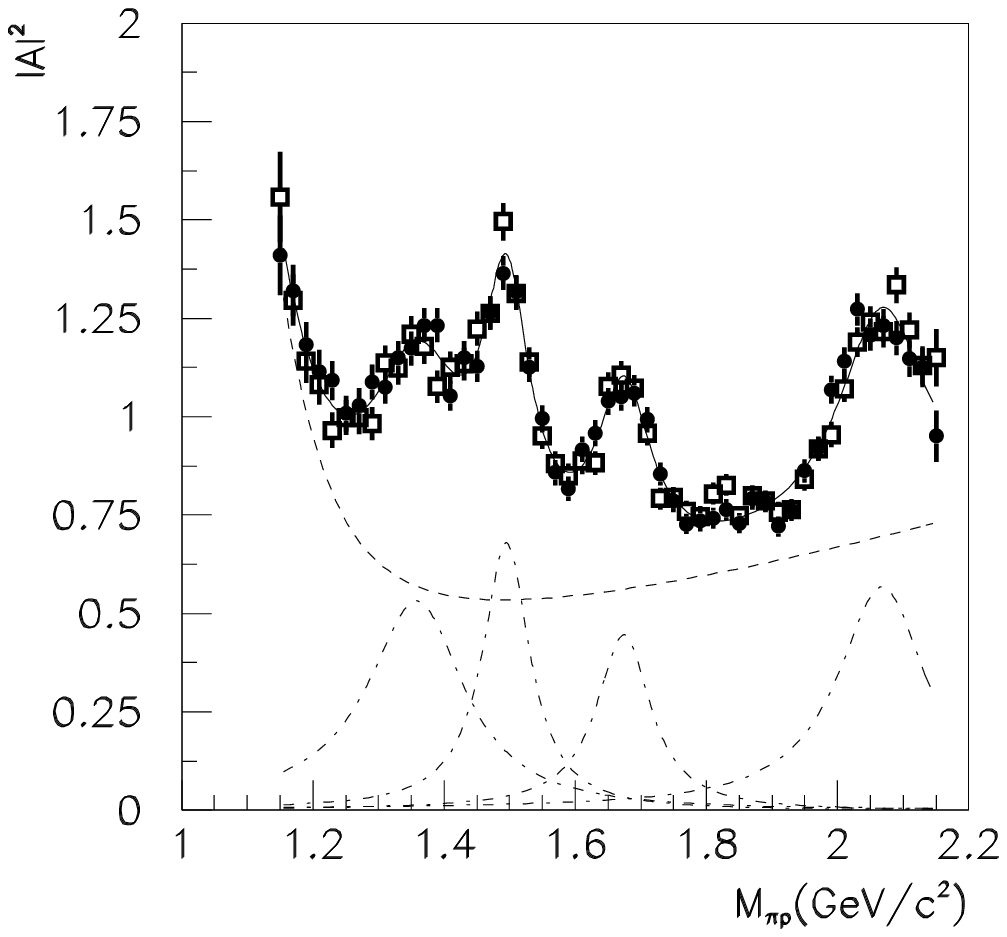, height=5.3cm,width=6.2cm}
\psfig{file=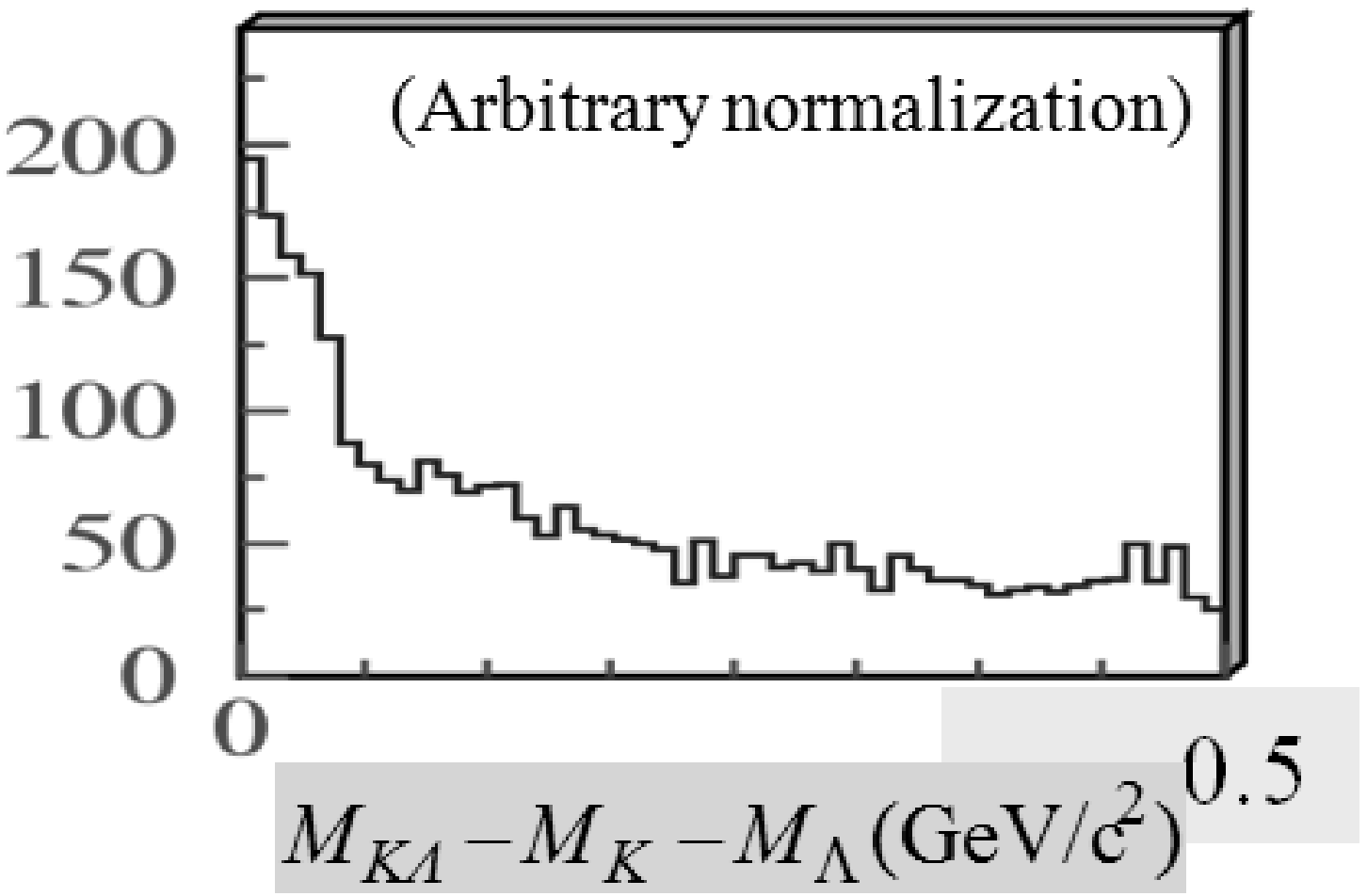,height=4.5cm,width=6.2cm} \caption{Invariant
mass spectrum divided by efficiency and phase space vs $n\pi$
invariant mass for $J/\psi\to\bar pn\pi^++c.c.^{25}$ and vs
$M_{K\Lambda}\!-\!M_K\!-\!M_\Lambda$ for $J/\psi\to
pK^-\bar\Lambda+c.c.^{26}$ .}
\label{fig:1}       
\end{figure}

Recently BES experiment at Beijing Electron-Positron Collider (BEPC)
has been producing very useful information on $N^*$ resonances
\cite{beseta,bespi,beska,weidh}. In $J/\psi\to\bar pp\eta$, as
expected, the $N^*(1535)$ gives the largest contribution
\cite{beseta}. In $J/\psi\to\bar pn\pi^++c.c.$ \cite{bespi}, a clear
peak containing $N^*(1535)$ contribution is observed around 1.5 GeV
in the $n\pi$ invariant mass spectrum as shown in Fig.~\ref{fig:1}
(left). In addition, a near-threshold enhancement due to
subthreshold nucleon pole contribution is clearly there. In
$J/\psi\to pK^-\bar\Lambda+c.c.$, a strong near-threshold
enhancement is observed for $K\Lambda$ invariant mass spectrum
\cite{beska} as shown in Fig.~\ref{fig:1} (right). The $K\Lambda$
threshold is 1609 MeV. The near-threshold enhancement is confirmed
by $J/\psi\to nK_S\bar\Lambda+c.c.$ \cite{weidh}. Since the mass
spectrum divided by efficiency and phase space peaks at threshold,
it is natural to assume it comes from the sub-threshold nearby
$N^*(1535)$ resonance decaying into $K\Lambda$ with relative S-wave.
Then from BES branching ratio results on $J/\psi\to\bar pp\eta$
\cite{beseta} and $J/\psi\to pK^-\bar\Lambda+c.c.$ \cite{beska}, the
ratio between effective coupling constants of $N^*(1535)$ to
$K\Lambda$ and $p\eta$ is deduced to be  \cite{liubc}
$$g_{N^*(1535)K\Lambda}/g_{N^*(1535)p\eta}
=1.3\pm 0.3 .$$

A previous well-known property of $N^*(1535)$ is its extraordinary
strong coupling to $\eta N$~\cite{pdg06}, which leads to a
suggestion that it is a quasibound ($K\Lambda-K\Sigma$)
state~\cite{kaiser}. This picture predicts large effective coupling
of $N^*(1535)$ to both $K\Lambda$ and $K\Sigma$ \cite{inoue}. While
the large $K\Lambda$ coupling seems confirmed here, the evidence for
large $K\Sigma$ coupling is still missing. An alternative picture
for the $N^*(1535)$ is that it contains large admixture of
$|[ud][us]\bar s>$ pentaquark component having $[ud]$, $[us]$ scalar
diquarks and $\bar s$ in the ground state \cite{liubc,zhusl}.  The
new picture expects large coupling to $K\Lambda$, but small coupling
to $K\Sigma$.

\begin{figure}[th]
\psfig{file=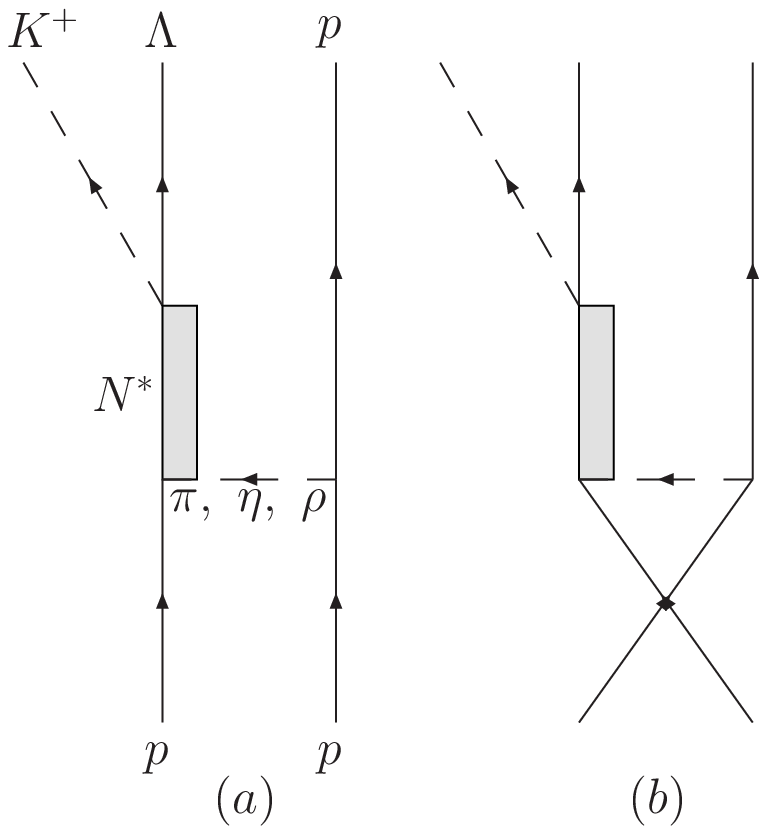,height=4.6cm,width=5.0cm} \hspace{0.8cm}
\psfig{file=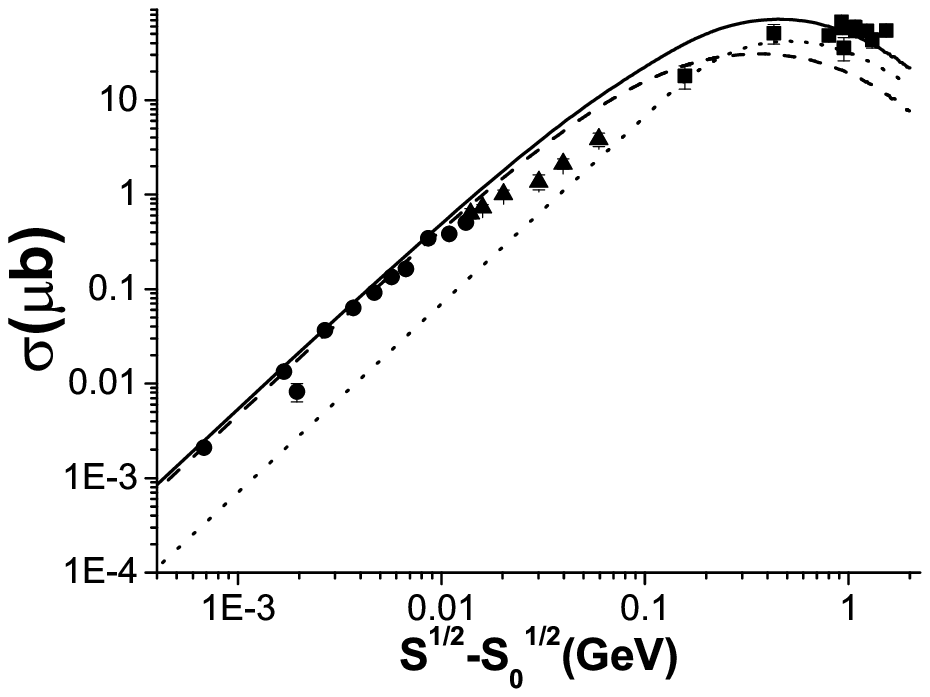,height=5.0cm,width=6.0cm} \caption{Feynman
diagrams (left) and calculated total cross section vs the excess
energy compared with data (right) for the reaction $pp \to pK^+
\Lambda^{19}$ .} \label{lkfd}
\end{figure}

No matter what picture is correct for the internal structure of the
$N^*(1535)$, its large coupling to $K\Lambda$ should also play a
role in other relevant reactions. Hence its possible contribution to
$pp \to pK^+\Lambda$ reaction is examined \cite{liubc} in the
effective Lagrangian framework with relevant Feynman diagrams as
plotted in Fig.~\ref{lkfd} (left) including t-channel exchange of
$\pi^0, \eta$ and $\rho^0$ mesons. The calculated results are shown
in Fig.~\ref{lkfd} (right). While the dotted line is taken from
Ref.~\cite{tsushima} which includes the contributions from
$N^*(1650)$, $N^*(1710)$ and $N^*(1720)$ resonances, the dashed line
is the contribution from the $N^*(1535)$ with its $K\Lambda$
coupling deduced from BES data, and the solid line is the sum. We
can see that the new results with the contribution from $N^*(1535)$
resonance reproduce the experiment data very well especially near
threshold.

\begin{figure}[thb]
\centerline{\psfig{file=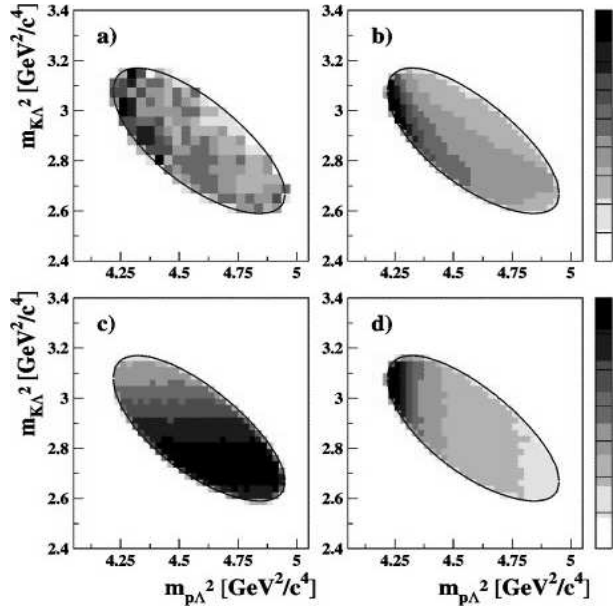,height=8cm,width=8cm}}
\caption{Dalitz plot$^{33}$ of the data at pbeam=2.85 GeV/c (a), in
comparison with the adjusted model of Sibirtsev (b), model
calculation only with the resonance part without FSI (c), and only
with p¦«-final-state interaction without resonances (d).
}\label{cosy-tof}
\end{figure}

The work \cite{liubc} got a comment from A.Sibirtsev et al.
\cite{liubc1c}. They pointed out that the work \cite{liubc} and
previous calculation \cite{tsushima} have not included the
$p\Lambda$ final state interaction (FSI). After including possible
$p\Lambda$ FSI, they can also reproduce the $pp \to pK^+\Lambda$
near-threshold total cross section data without inclusion of the
$N^*(1535)$ contribution. However, recent COSY-TOF data on Dalitz
plot \cite{cosytof} clearly demonstrated that besides the $p\Lambda$
near-threshold enhancement due to $p\Lambda$ FSI there is also a
$K\Lambda$ near-threshold enhancement as shown in
Fig.~\ref{cosy-tof}(a) which cannot be reproduced by the Sibirtsev
model simulation without including the $N^*(1535)$
(Fig.~\ref{cosy-tof}(b). Obviously both $p\Lambda$ FSI and
$N^*(1535)$ contribution are necessary. With $p\Lambda$ FSI
included, the large $K\Lambda$ coupling deduced from BES data for
$N^*(1535)$ is still found compatible with $pp \to pK^+\Lambda$ data
\cite{liubc2}.

There are also indications for the large $g_{N^*(1535)K\Lambda}$
from partial wave analysis of $\gamma p\to K\Lambda$ reactions
\cite{Lee} and evidence for large $g_{N^*(1535)N\eta^\prime}$
coupling from $\gamma p \to p\eta^\prime$ reaction at CLAS
\cite{etap}.

\section{Study on $pp \to pp\phi$ reaction}

\begin{figure}[th]
\psfig{file=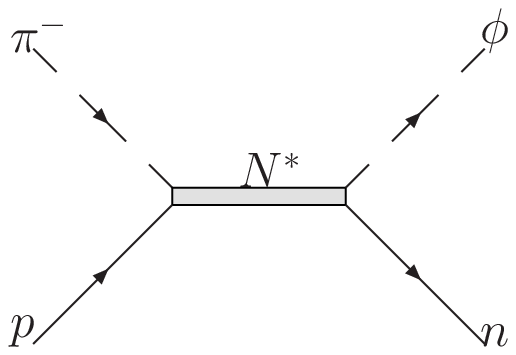,height=4cm,width=5cm}
\psfig{file=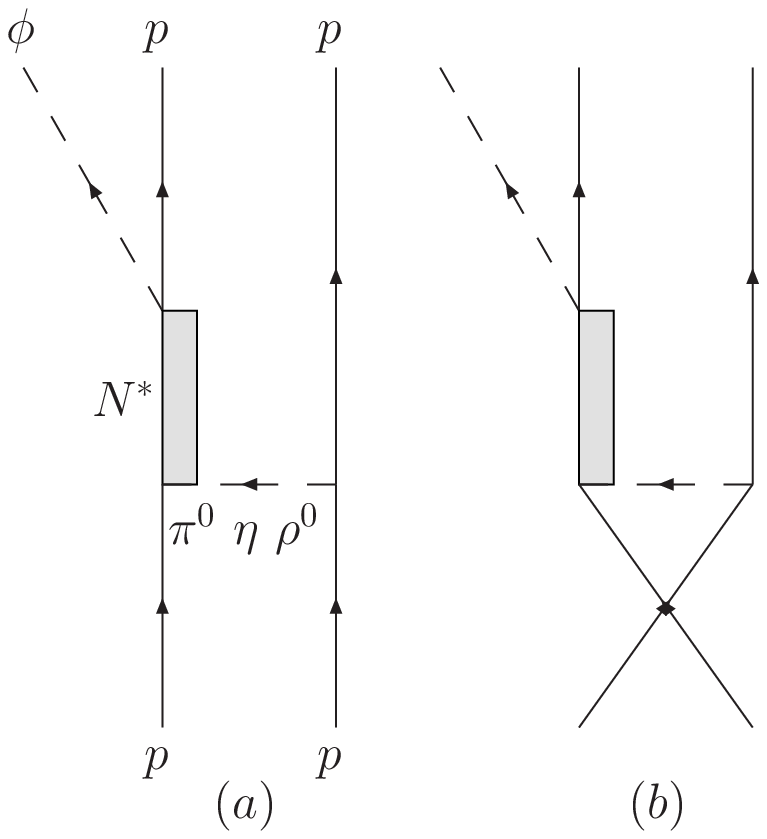,height=5cm,width=7cm} \caption{Feynman
diagrams for $\pi^- p \to n\phi$ and $pp \to pp \phi$ reactions.}
\label{pipfd}
\end{figure}

In the naive quark model, the nucleon and nucleon resonances have no
strangeness contents, whereas the $\phi$ meson is an ideally mixed
pure $s \bar{s}$ state. From the point of view of the naive quark
model the $pp \to pp \phi$ reaction involves disconnected quark
lines and is an Okubo-Zweig-Iizuka (OZI) rule suppressed process.
The study of  $\phi$ meson production in nucleon-nucleon reactions
may provide information on the strangeness degrees of freedom in the
nucleon or nucleon resonances and is of importance both
experimentally and theoretically. Since $N^*(1535)$ resonance has
strong coupling to $N\eta$,  $K\Lambda$ and maybe also
$N\eta^\prime$, there may be a significant $s\bar{s}$ configuration
in the quark wave function of the $N^*(1535)$ resonance. So the
$N^*(1535)$ resonance may also have a significant coupling to the
$\phi N$ channel. Assuming that the productions of the $\phi$ meson
in $pp$ and $\pi^- p$ collisions are predominantly through the
excitation and decay of the sub-$\phi N$-threshold $N^*(1535)$
resonance, we calculated the $pp \to pp \phi$ and $\pi^- p \to n
\phi$ reactions in the framework of an effective lagrangian
approach~\cite{xiephi}. A Lorentz covariant orbital-spin (L-S)
scheme~\cite{zouls} is used for the effective interaction vertices
involving the $N^*(1535)$ resonance. The relevant Feynman diagrams
considered in our computation are shown in Fig.~\ref{pipfd}.

\begin{figure}[th]
\psfig{file=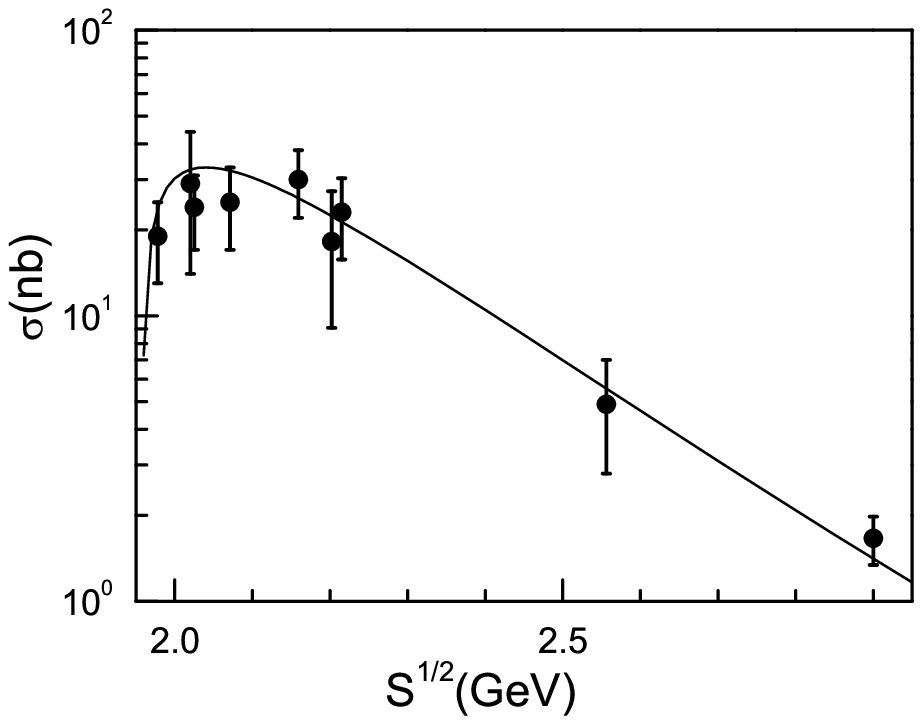,height=5.5cm,width=6.2cm}
\psfig{file=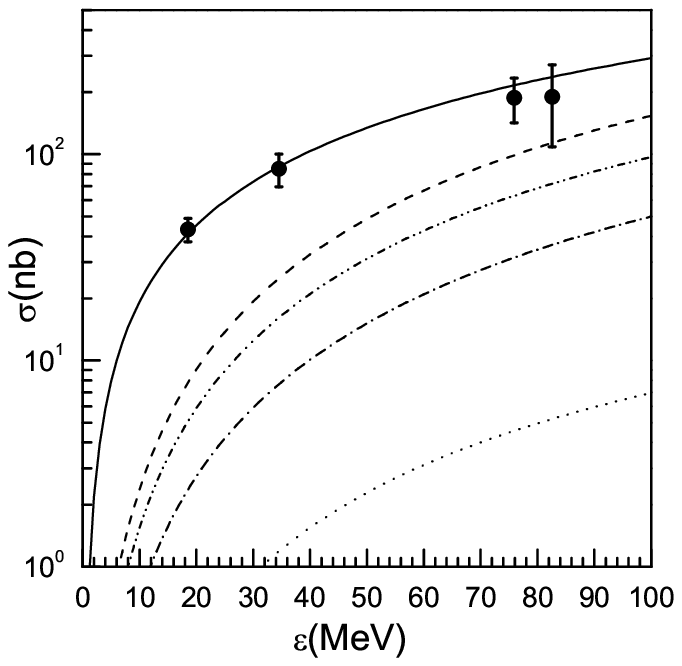,height=5.5cm,width=6.2cm} \caption{The total
cross section vs the C.M. energy for the $\pi^-p \to n\phi$ reaction
(left) and vs excess energy for the $pp \to pp\phi$ reaction (right)
compared with the relevant experimental data.} \label{piptcs}
\end{figure}

There is no information for the coupling constant of the
$N^*(1535)N\phi$ vertex. We determine it from the $\pi^- p \to
n\phi$ reaction. We calculated the total cross section of the
reaction based on $s$-channel $N^*(1535)$ excitation since
contributions from the $u$-channel $N^*(1535)$ excitation and
$t$-channel $\rho$-meson exchange are also checked and are found to
be negligible. By adjusting the $N^*(1535)N\phi$ coupling constant,
we can compare the theoretical results with the experimental data.
Theoretical results with $g_{N^*(1535)N\phi}$ = 0.13 are compared
with the experimental data by the solid curve in Fig.~\ref{piptcs}
(left). We find an excellent agreement between our results and the
experimental data.

Then we calculated the total cross section of the $pp \to pp\phi$
reaction with $\pi^0, \eta$ and $\rho^0$ mesons exchange for
$N^*(1535)$ excitation. The numerical results are shown in
Fig.~\ref{piptcs} (right) together with the experimental
data~\cite{balestra,hartman}. The double dotted-dashed, dotted,
dashed-dotted and dashed curves stand for contributions from
$\pi^0$, $\eta$, $\rho^0$-meson exchanges and their simple sum,
respectively. To show the effect from the $pp$ final state
interaction (FSI), we give the results with the $^1S_0$ $pp$ FSI by
solid line in the figure. One can see that the contribution from the
$\pi$ meson exchange is dominant to the $pp \to pp \phi$ reaction in
our model. The $\rho$ meson exchange has a significant contribution
to this reaction, while the contribution from the $\eta$ meson
exchange is negligible.

In our calculation we only include the contribution of the
$N^*(1535)$ in the intermediate state. In previous calculations
\cite{sibiepja06,titov,nakaphi,kap}, the $\pi p\to\phi N$ through
t-channel $\rho$ exchange and/or sub-threshold nucleon pole
contributions are assumed to be dominant. However these
contributions are very sensitive to the choice of off-shell form
factors for the t-channel $\rho$ exchange and the $g_{NN\phi}$
couplings and can be reduced by orders of magnitude within the
uncertainties of these ingredients. Considering the ample evidence
for large coupling of the $N^*(1535)$ to the strangeness
\cite{liubc,inoue,Lee,Riska} and the $N^*(1535)$ resonance is closer
than the nucleon pole to the $\phi N$ threshold, it is more likely
that the $N^*(1535)$ plays dominant role for near threshold $\phi$
production from $\pi p$ and $pp$ collisions instead of the nucleon
pole or the OZI suppressed $\phi\rho\pi$ coupling. Our calculation
with the $N^*(1535)$ domination reproduces energy dependence of the
$\pi^-p\to\phi n$ and $pp\to pp\phi$ cross sections better than
previous calculations. The significant coupling of the $N^*(1535)$
resonance to $N \phi$ may be the real origin of the significant
enhancement of the $\phi$ production from $\pi p$ and $pp$ reactions
over the naive OZI-rule predictions. This makes it difficult to
extract the properties of the strangeness in the nucleon from these
reactions proposed by J.Ellis et al \cite{Ellis}. There are also
some suggestions \cite{gao,zhang} for possible existence of an
$N\phi$ bound state just below the $N\phi$ threshold. However,
contribution of such bound state with width less than 100 MeV will
give a much sharper dropping structure for the $\pi^-p\to\phi n$
cross section at energies near threshold. If such $N\phi$ bound
state does exist, it should have weak coupling to $\pi N$ and only
gives small contribution to the $\pi^-p\to\phi n$ reaction.

However, we cannot exclude alternative solutions with significant
contributions from $N^*(1900)$ or $N^*(1650)$ although there are
some arguments favoring the solution with the dominant $N^*(1535)$
contribution. For a better understanding of the dynamics of these
reactions, more experimental data at other excess energies with
Dalitz plots and angular distributions are desired.

\section{Study on $pp \to n K^+\Sigma^+$ reaction}

The spectrum of isospin 3/2 $\Delta^{++*}$ resonances is of special
interest since it is the most experimentally accessible system
composed of 3 identical valence quarks. However, our knowledge on
these resonances mainly comes from old $\pi N$ experiments and is
still very poor~\cite{pdg06}. A possible new excellent source for
studying $\Delta^{++*}$ resonances is $pp \to nK^+\Sigma^+$
reaction, which has a special advantage for absence of complication
caused by $N^*$ contribution because of the isospin and charge
conversation.

\begin{figure}[th]
\centerline{\psfig{file=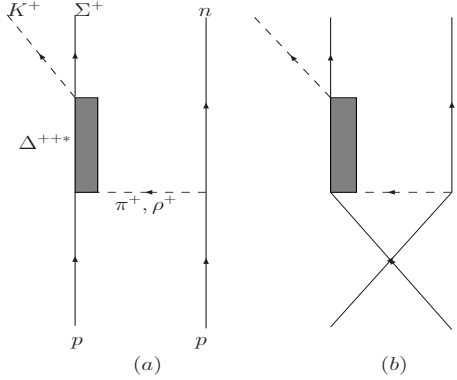,height=5.0cm,width=6.0cm}}
\vspace*{8pt} \caption{Feynman diagrams for the reaction $pp \to
nK^+ \Sigma^+$.} \label{nkfd}
\end{figure}

At present, little is known about the $pp\to nK^+\Sigma^+$ reaction.
Experimentally there are only a few data points about its total
cross section versus energy~\cite{rozek,baldi}. Theoretically a
resonance model with an effective intermediate $\Delta^{++*}(1920)$
resonance~\cite{tsushima99} and the J\"{u}lich  meson exchange
model~\cite{gaspar} reproduce the old data at higher beam
energy~\cite{baldi} quite well, but their predictions for the cross
sections close to threshold fail by order of magnitude compared with
very recent COSY-11 measurement~\cite{rozek}.

Recently this reaction was restudied \cite{xieplb}. For the $pp \to
n K^+\Sigma^+$, the basic Feynman diagrams are depicted in
Fig.~\ref{nkfd}. Besides the ingredients considered in previous
calculations~\cite{tsushima99,gaspar,shyam06}, the
sub-$K\Sigma$-threshold $\Delta^{++*}(1620)$ resonance is added by
taking into account both $\pi^+$ and $\rho^+$ mesons exchange.

\begin{figure}[th]
\centerline{\psfig{file=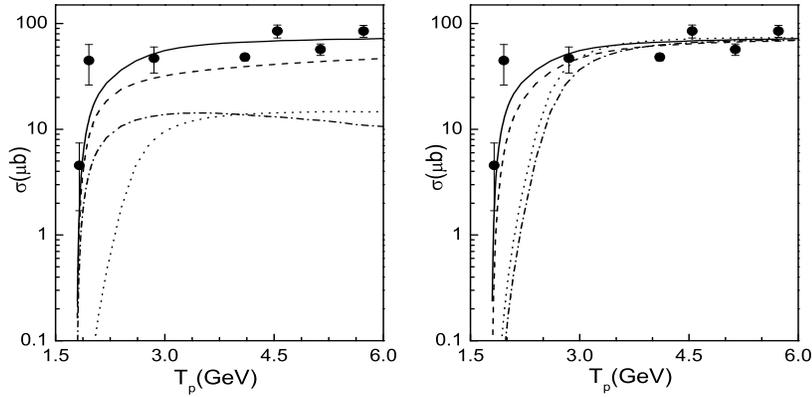,height=6.5cm,width=12.0cm}}
\vspace*{8pt} \caption{The total cross section vs $T_P$ for the $pp
\to nK^+\Sigma^+$ reaction compared with the relevant experimental
data.} \label{nktcs}
\end{figure}

The numerical results are shown in Fig.~\ref{nktcs} together with
the experimental data~\cite{rozek,baldi} for comparison. In the left
of Fig.~\ref{nktcs}, contributions from $\Delta^*(1620)(\pi^+$
exchange), $\Delta^*(1620)(\rho^+$ exchange) and
$\Delta^*(1920)(\pi^+$ exchange) are shown separately by dot-dashed,
dashed and dotted curves, respectively. The contribution from the
$\Delta^*(1620)$ production by the $\rho^+$ exchange is found to be
very important for the whole energy range, in particular, for the
two lowest data points close to the threshold. This gives a natural
source for the serious underestimation of the near-threshold cross
sections by previous calculations \cite{tsushima99,gaspar,shyam06},
which have neglected either $\Delta^*(1620)$ resonance contribution
\cite{tsushima99,shyam06} or $\rho^+$ exchange contribution
\cite{gaspar}. The solid curve in the figure is the incoherent sum
of the three contributions and reproduces the experimental data
quite well.

To show the effect from the $n$-$\Sigma^+$ final state interaction
(FSI), we give the result without including the FSI factor by the
dashed curve in the right figure of Fig.~\ref{nktcs}. Comparing
dashed curve with the solid curve which includes the FSI factor, we
find that the FSI enhances the total cross section by a factor of
about 3 for the two lowest data points. So the FSI is indeed making
a significant effect at energies close to threshold. But it does not
change the basic shape of the curve very much. In previous
calculations \cite{tsushima99,shyam06}, only $\Delta^*(1920)$
contribution is considered with a free scaling parameter to fit the
data. In Fig.~\ref{nktcs} (right), we also show the results from
only $\Delta^*(1920)(\pi^+ $ exchange) scaled by a factor 5 for
comparison. It reproduces the data for T$_\text P$ above 2.8 GeV
quite well, but underestimates the two lowest data points by orders
of magnitude no matter whether including the FSI (dotted curve) or
not (dot-dashed curve).

Meanwhile the extra-ordinary large coupling of the
$\Delta^{*}(1620)$ to $\rho N$ obtained from the $\pi^+ p\to
N\pi\pi$ \cite{pdg06,dytman} seems confirmed by the new study
\cite{xieplb} of the strong near-threshold enhancement of $pp \to
nK^+\Sigma^+$ cross section. Does the $\Delta^{*}(1620)$ contain a
large $\rho N$ molecular component or relate to some $\rho N$
dynamical generated state? If so, where to search for its SU(3)
decuplet partners? Sarkar et al.~\cite{sarkar} have studied baryonic
resonances from baryon decuplet and psudoscalar meson octet
interaction. It would be of interests to study baryonic resonances
from baryon octet and vector meson octet interaction. In fact, from
PDG compilation \cite{pdg06} of baryon resonances, there are already
some indications for a vector-meson-baryon SU(3) decuplet. While the
$\Delta^{*}(1620)1/2^-$ is about 85 MeV below the $N\rho$ threshold,
there is a $\Sigma^*(1750)1/2^-$ about 70 MeV below the $NK^*$
threshold and there is a $\Xi^*(1950)?^?$ about 60 MeV below the
$\Lambda K^*$ threshold. If these resonances are indeed the members
of the $1/2^-$ SU(3) decuplet vector-meson-baryon S-wave states, we
would expect also a $\Omega^* 1/2^-$ resonance around 2160 MeV. All
these baryon resonances can be searched for in high statistic data
on relevant channels from vector charmonium decays by upcoming BES3
experiments in near future.

\section{Summary}

In this work, we reviewed the important role played by subthreshold
$N^*(1535)$ and $\Delta^*(1620)$ resonances to $pp \to pK^+\Lambda$,
$pp \to pp \phi$ and $pp \to nK^+\Sigma^+$ reactions. While
$N^*(1535)$ resonance plays a dominant role for the near-threshold
total cross sections of $pp \to pK^+\Lambda$,  $pp \to pp \phi$ and
$\pi^- p \to n \phi$ reactions, the $\Delta^*(1620)$ resonance plays
a dominant role in $pp \to nK^+\Sigma^+$ reaction. They are crucial
ingredients for reproducing data of the strangeness production in
$pp$ collisions

The results have many important implications:

(1) Since the $pp \to pK^+\Lambda$, $pp \to nK^+\Sigma^+$ and $pp
\to pp \phi$ reactions are the basic inputs for the strangeness
production in heavy ion collisions \cite{rafel,liplb}, the inclusion
of the sub-threshold $N^*(1535)$ and $\Delta^*(1620)$ resonances
contributions may be essential for such studies.

(2) They give new examples that sub-threshold resonances can make
extremely important contributions and should not be simply ignored.
Many calculations were used to consider only the resonances above
threshold, such as previous calculations \cite{tsushima99,shyam06}
for $pp \to pK^+\Lambda$ and $pp \to nK^+\Sigma^+$ reactions. There
are several more examples from $J/\psi$ decays showing the
importance of contribution from sub-threshold particles, such
sub-$\pi N$-threshold nucleon pole contribution in $J/\psi\to \bar
pn\pi^+$ \cite{beseta,liang}, sub-$K\bar K$-threshold contribution
in $J/\psi\to K\bar K\pi$ and sub-$\omega\pi$-threshold contribution
in $J/\psi\to \omega\pi\pi$ \cite{wufq}.

(3) The t-channel $\rho$ exchange may play an important role for
many meson production processes in $pp$ collisions and should not be
ignored.

(4) While the classical 3q constituent quark model works well in
reproducing properties of baryons in the spatial ground states, the
study of $1/2^-$ baryons seems telling us that the $\bar qqqqq$ in
S-state is more favorable than $qqq$ with $L=1$. In other words, for
excited baryons, the excitation energy for a spatial excitation
could be larger than to drag out a $q\bar q$ pair from gluon field.
Whether the $\bar qqqqq$ components are in penta-quark configuration
or meson-baryon configuration depends on the strength of relevant
diquark or meson-baryon correlations. For $N^*(1535)$ and its
$1/2^-$ SU(3) nonet partners, the diquark cluster picture for the
penta-quark configuration gives a natural explanation for the
longstanding mass-reverse problem of $N^*(1535)$, $N^*(1440)$ and
$\Lambda^*(1405)$ resonances as well as the unusual decay pattern of
the $N^*(1535)$ resonance. Its predictions of the existence of an
additional $\Lambda^*~1/2^-$ around 1570 MeV, a triplet
$\Sigma^*~1/2^-$ around 1360 MeV and a doublet $\Xi^*~1/2^-$ around
1520 MeV \cite{zhusl} could be examined by forth coming experiments
at BEPC2, CEBAF, JPARC etc.. For $\Delta^{*++}(1620)$ and its
$1/2^-$ SU(3) decuplet partners, their SU(3) quantum numbers do not
allow them to be formed from two good scalar diquarks plus a $\bar
q$. Then their $\bar qqqqq$ components would be mainly in the
meson-baryon configuration. This picture can be also examined by
forth coming experiments.

\section*{Acknowledgements}

We would like to thank B.C.Liu and H.C.Chiang for collaborations on
some relevant issues reviewed here. This work is partly supported by
the National Natural Science Foundation of China under grants Nos.
10435080, 10521003 and by the Chinese Academy of Sciences under
project No. KJCX3-SYW-N2.

\end{document}